\documentclass[aps,prb,twocolumn,superscriptaddress]{revtex4-1}
\usepackage{amssymb}
\usepackage{graphicx}
\usepackage{dcolumn}
\usepackage{bm}
\usepackage{amsmath}
\usepackage{epstopdf}
\usepackage{xcolor}
\usepackage{float}
\usepackage[english]{babel}
\usepackage[colorlinks,linkcolor=magenta,citecolor=blue,urlcolor=blue]{hyperref}
\begin{document}

\title{Fate of Majorana zero modes, critical states and non-conventional real-complex transition in non-Hermitian quasiperiodic lattices}

\author{Tong Liu}
\thanks{t6tong@njupt.edu.cn}
\affiliation{Department of Applied Physics, School of Science, Nanjing University of Posts and Telecommunications, Nanjing 210003, China}
\author{Shujie Cheng}
\thanks{2818917376@qq.com}
\affiliation{Department of Physics, Zhejiang Normal University, Jinhua 321004, China}
\author{Hao Guo}
\thanks{guohao.ph@seu.edu.cn}
\affiliation{Department of Physics, Southeast University, Nanjing 211189, China}
\author{Gao Xianlong}
\thanks{gaoxl@zjnu.edu.cn}
\affiliation{Department of Physics, Zhejiang Normal University, Jinhua 321004, China}

\date{\today}

\begin{abstract}
  We study a one-dimensional $p$-wave superconductor subject to non-Hermitian quasiperiodic potentials. Although the existence of the non-Hermiticity, the Majorana zero mode is still robust against the disorder perturbation. The analytic topological phase boundary is verified by calculating the energy gap closing point and the topological invariant. Furthermore, we investigate the localized properties of this model, revealing that the topological phase transition is accompanied with the Anderson localization phase transition,  and a wide critical phase emerges with amplitude increments of the non-Hermitian quasiperiodic potentials. Finally, we numerically uncover a non-conventional real-complex transition of the energy spectrum, which is different from the conventional $\mathcal{PT}$ symmetric transition.

\end{abstract}

\maketitle

\section{Introduction}
\label{n1}
The discovery of Anderson localization~\cite{Anderson} is giving guidance to
understand how the disorder affects the mobility of carriers through the spatial distribution of the wave function beyond the framework
of the conventional band theory. After half a century, the Anderson localization phenomena were observed in a ultracold atom experiment
with correlated disordered potentials \cite{disorder} and incommensurate/quasiperiodic potentials \cite{quasiperiodic}.
Nowadays, Anderson localization has been one of the important and highly-explored
research subjects in condensed matter physics~\cite{MBL_1,MBL_2,MBL_3,MBL_4}. In one-dimensional systems, many researches show that even though there exist particle interactions, random disordered or incommensurate disordered external potentials can form the many-body localization \cite{MBL_5,MBL_6,MBL_7,MBL_8,MBL_9,MBL_10}, which is the many-body version of Anderson localization.

A paradigmatic model to understand the Anderson localization is the Aubry-Andr\'{e}-Harper (AAH) model \cite{AAH_1,AAH_2}, in which the increased strength of the incommensurate potential leads to a localized transition. In some generalized AAH models, the rich localization phenomena can be observed \cite{extended_1,extended_2,extended_3,extended_4,extended_5}.
Another interesting aspect of generalized AAH models is the presence of the mobility edge, which characterizes the separation between the
extended and the localized region in terms of energy~\cite{extended_5,mobility_1,mobility_2,mobility_3,mobility_4,mobility_5,mobility_6}.

It should be noted that the one-dimensional AAH model can be understood as the projection of the two-dimensional Hofstadter model in the one-dimensional direction \cite{AAH_1,proj_1,proj_2}, which supports topologically protected edge states localized at the boundary, similar to the edge
states in quantum Hall insulators \cite{QHI_1,QHI_2}. Consequently, the topological properties of one-dimensional quasicrystals have been
gradually excavated according to this projection \cite{tp_AAH_1,tp_AAH_2,tp_AAH_3,tp_AAH_4,tp_AAH_5}.
In topology physics, the one-dimensional $p$-wave superconductor chain is another important paradigm~\cite{K1,Zh2,K3,I4}. A key feature of the one-dimensional $p$-wave superconductor is that it hosts topologically protected Majorana zero mode (MZM)~\cite{S5,K6,Lu7}, which promise a platform for the error-free quantum computation since the qubits are immune to the weakly disordered perturbation~\cite{P8,MZM_2}.
Thus, the interplay of disorder and topology in one-dimensional quasiperiodic lattices with $p$-wave superconducting pairing attracted much research interest. In a previous research, Cai et.al. uncovered that the topological phase transition is accompanied by the Anderson localization phase transition in a Hermitian quasiperiodic chain with $p$-wave superconducting pairing \cite{cai}. Further research showed that there is a critical phase in the topologically non-trivial region \cite{proj_2}.

However, when the non-Hermitian and quasiperiodic potentials are both considered, how does the topological region change compared to the Hermitian case, does the critical phase still exist, and does there exist real-complex transition of eigenenergies, even if the system is not $\mathcal{PT}$ symmetric? In this paper, we are devoted to answer these questions. Specifically, we deduce the phase boundary of topological phase
transition analytically and verify it from the energy gap, the spatial distributions of the MZMs and the topological invariant. In addition,
we clarify the extended, critical and localized regions of this model by use of the fractal theory. Finally, we uncover a real-complex transition
of the system without $\mathcal{PT}$ symmetry, and give a conclusion that the extended phase corresponds to the region where eigenenergies are totally real, whereas the critical and localized phase correspond to the region where eigenenergies are complex.

The arrangement of the rest paper is as follows: Sec.~\ref{n2} describes the Hamiltonian of the one-dimensional $p$-wave superconductor subject to the non-Hermitian quasiperiodic potentials and gives the definition of the inverse participation ratio; Sec.~\ref{n3} discusses the fate of Majorana zero modes and the topological phase transition; Sec.~\ref{n4} discusses Anderson localization phase transition and the critical phase; Sec.~\ref{n5} discusses the non-conventional real-complex transition of the energy spectrum and presents the total phase diagram of the system; we make a summary in Sec.~\ref{n6}.

\section{Model and Hamiltonian}
\label{n2}
We consider the one-dimensional $p$-wave superconductor subject to the non-Hermitian quasiperiodic potentials, which is described by the following Hamiltonian
\begin{equation}\label{eq1}
    \hat H=\sum_{n=1}^{L-1}(-t \hat{c}_{n}^{\dag } \hat{c}_{n+1}+ \Delta \hat{c}_{n} \hat{c}_{n+1}+ H.c.)+\sum_{n=1}^{L}V_{n} \hat{c}_{n}^{\dag } \hat{c}_{n},
\end{equation}
where $\hat{c}^\dagger_n$ ($\hat{c}_n$) is the fermion creation (annihilation) operator, and $L$ is the total number of sites. Here
the nearest-neighbor hopping amplitude $t$ and the $p$-wave pairing amplitude $\Delta$ are real constants, and $V_{n}=V e^{i 2\pi\alpha n}$
is the non-Hermitian quasiperiodic potential. A typical choice for parameter $\alpha$ is $\alpha=(\sqrt{5}-1)/2$. For computational convenience, $t = 1$ is set as the energy unit. In the topological classification,
this model belongs to the BDI class \cite{classify} and it does not preserve $\mathcal{PT}$-symmetry \cite{PT}. When $\Delta$ is equal to zero, this model reduces to the non-Hermitian AAH model \cite{mobility_2}, where the localized transition and the topological properties are well understood.
When $\alpha=0$, this Hamiltonian describes the Kitaev model, where there are topologically protected MZMs \cite{K3,MZM_2}. When the imaginary part of the non-Hermitian
potential is omitted, the model reduces to the Hermitian non-Abelian AAH model \cite{proj_2,cai}, in which the topological phase transition and the Anderson localization transition is well studied.

The Hamiltonian~(\ref{eq1}) can be diagonalized by using the Bogoliubov-de Gennes (BdG) transformation:
\begin{equation}\label{eq2}
\hat{\chi} _{m}^{\dag } = \sum_{n=1}^{L}[u _{m,n} \hat{c}_{n}^{\dag } +v _{m,n} \hat{c}_{n}],
\end{equation}
where $L$ denotes the total number of sites, $n$ is the site index, and $u_{m,n}$, $v_{m,n}$ are the two components of wave functions.
It is widely known that the particle-hole symmetry is preserved \cite{K3}. Under this transformation, the BdG equations can be
expressed as
%Hence theHamiltonian is diagonalized as $H=\sum_{m=1}^{L}E _{m}(\hat{\chi}_{m}^{\dag }\hat{\chi} _{m}-\frac{1}{2})$ where $E_{m}$ is the eigenenergy of the Hamiltonian.

\begin{eqnarray}
 \left(
\begin{array}{cc}
\hat{M} & \hat{\Delta} \\
-\hat{\Delta} & -\hat{M}%
\end{array}
\right)
 \left(
\begin{array}{c}
u _{m} \\
v _{m}%
\end{array}%
\right) =
E _{m} \left(
\begin{array}{c}
u_m \\
v_{m}%
\end{array}
\right), \label{BDG}
\end{eqnarray}
where $ \hat{M}_{ij} = -t (\delta_{j,i+1} + \delta_{j,i-1}) +V_{i}  \delta_{ji}$, $\hat{\Delta}_{ij} = - \Delta(\delta_{j,i+1}-\delta_{j,i-1})$,
$u_m^T=(u_{m,1},\cdots,u_{m,L})$ and $v_m^T=(v_{m,1},\cdots,v_{m,L})$, $E_m$ is the complex eigenenergy, indexed according to its real part
${\rm Re}(E_m)$ and arranged in ascending order with $m$ being the energy level index.

By numerically solving Eq.~(\ref{BDG}), we can obtain the energy spectrum of the system and the components $u_{m,n}$ and $v_{m,n}$
of the wave functions. The inverse participation ratio (IPR) is usually used to study the localization-delocalization transition \cite{AAH_1,mobility_2,proj_2,cai}.
For any given normalized wave function, the corresponding IPR is defined as
\begin{equation}
{\rm{IPR}}_m =\frac{\sum^{L}_{n=1}\left(\left|u_{m,n}\right|^{4}+\left|v_{m,n}\right|^{4}\right)}{\left[\sum^{L}_{n=1}(\left|u_{m,n}\right|^{2}+\left|v_{m,n}\right|^{2})\right]^{2}},
\end{equation}
which measures the inverse of the number of sites being occupied by particles. It is well known that the IPR of an extended state scales
like $L^{-1}$ which approaches zero in the thermodynamic limit. However, for a localized state, since only finite number of sites are
occupied, the IPR is finite even in the thermodynamic limit. The mean of IPR over all the $2L$ eigenstates is dubbed the MIPR which is
expressed as
\begin{equation}
{\rm{MIPR}}=\frac{1}{2L}\sum_{m=1}^{2L}{\rm{IPR}}_{m}.
\end{equation}

\section{Fate of Majorana zero modes}
\label{n3}

\begin{figure}%[H]
	\centering
	\includegraphics[width=0.5\textwidth]{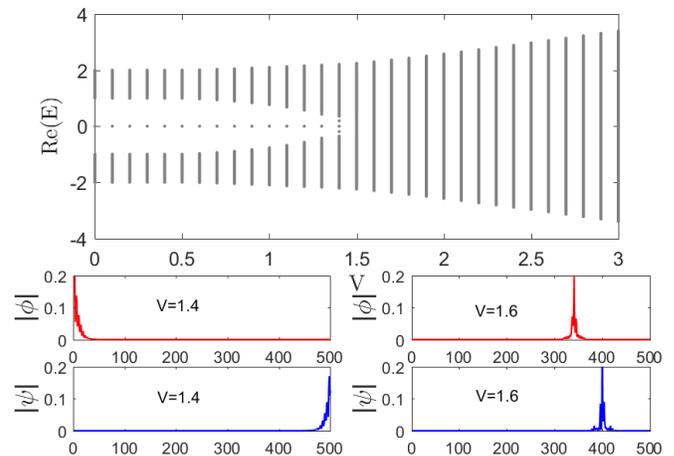}\\
	\caption{(Color online) Top panel: The real part of eigenvalues of Eq.~(\ref{eq1}) as a function of $V$ under OBC. Definitely,
there are stable MZMs when $V<1+\Delta$. As the value of $V$ continuously increases, the MZM eventually vanishes, and the
phase transition point is roughly located at $V_{c}=1+\Delta$. Bottom panel: Spatial distributions of $\phi$ and $\psi$ for
the lowest excitation modes with $V=1.4$ in the left and with $V=1.6$ in the right. Obviously, when $V=1.4$, $\phi$ and $\psi$
are symmetrically distributed at ends of the chain, which indicates the system is in topological phase, whereas they are located inside of the
chain when $V=1.6$. Other involved parameters are $\Delta=0.5$ and $L=500$.}
	\label{f1}
\end{figure}

In this part, we will study the fate of the MZMs and the topological phase transition. 
The top panel in Fig.~\ref{f1} shows the real part of the energy spectrum of Eq.~(\ref{eq1}) as a function of the non-Hermitian potential strength $V$
under the open boundary condition (OBC), with the parameters $\Delta=0.5$ and $L=500$. As shown in the figure, there are stable MZMs when
$V<1+\Delta$. However, when $V$ is larger than the critical value $V_{c}$, MZMs annihilate and then enter into the bulk of the system. Hence, the systems will
undergo a topological non-trivial to trivial phase transition as $V$ increases, and the visible phase transition point is about $V_{c}=1+\Delta$.
Similar to the previous works \cite{proj_2,K3,MZM_2,cai}, MZMs in our system are still localized at ends of the system. To understand the Majorana edge state deeply, we have to introduce the Majorana operators, namely $\lambda^{A}_{n}=\hat{c}^{\dag}_{n}+\hat{c}_{n}$
and $\lambda^{B}_{n}=i(\hat{c}^{\dag}_{n}-\hat{c}_{n})$, which obey the relations $(\lambda^{\beta}_{n})^\dag=\lambda^{\beta}_{n}$ and $\left\{\lambda^{\beta}_{n},\lambda^{\beta'}_{n'}\right\}=2\delta_{nn'}\delta_{\beta\beta'}$, with $\beta,~\beta' \in \left\{A,~B\right\}$.
Accordingly, in the Majorana picture, the quasi-particle operator in Eq.~(\ref{eq2}) can be rewritten as
\begin{equation}
\hat{\chi} _{m}^{\dag } = \frac{1}{2}\sum_{n=1}^{L}[\phi_{m,n} \lambda^{A}_{n}-i\psi_{m,n}\lambda^{B}_{n}],
\end{equation}
in which $\phi_{m,n}=(u_{m,n}+v_{m,n})$ and $\psi_{m,n}=(u_{m,n}-v_{m,n})$.

The bottom panel of Fig.~\ref{f1} plots the spatial distributions of $\phi$ and $\psi$ for the lowest excitation mode \cite{cai,t6tong} under OBC,
with $\Delta=0.5$ and $V=1.4$ (left bottom panel) and $V=1.6$ (right bottom panel). When $V=1.4$, the lowest excitation mode is just the MZM. As
the corresponding figures show, the Majorana edge states $\phi$ and $\psi$ are localized at ends of the system, presenting the chiral symmetry.
On the contrary, when $V=1.6$, the lowest excitation mode is no longer the MZM. As a result, the visible distributions of $\phi$ and $\psi$ in
the right bottom panel are located inside the bulk of the system. Therefore, only if $V$ is less than $V_{c}$, the system is topologically
non-trivial and supports the MZM.

Due to the bulk-edge correspondence, the topological properties of non-Hermitian systems are generally protected by the real gaps \cite{MZM_2,real_gap}.
In other words, topological phase transition occurs with the gap closing. Before we clarify the relationship between the topologically non-trivial
phase and the real gap, we first deduce the topological phase transition point $V_{c}$. Under the periodic boundary condition (PBC), the Hamiltonian
in Eq.~(\ref{eq1}) can be rewritten as
\begin{equation}
\hat{H}=\sum_{nn'}\left[\hat{c}^{\dag}_{n}M_{nn'}\hat{c}_{n'}+\frac{1}{2}\left(\hat{c}^{\dag}_{n}N_{nn'}\hat{c}^{\dag}_{n'}+h.c.\right)\right],
\end{equation}
where M is a Hermitian matrix and N is an antisymmetric matrix, respectively expressed as
\begin{equation}
M=\left(
\begin{array}{cccc}
  V_1 & -t &   \cdots & -t \\
  -t & V_2 &   &  \\
  \vdots &    & \ddots & -t \\
  -t &    & -t & V_L
\end{array}
\right),
N=\left(
\begin{array}{cccc}
  0 & -\Delta &   \cdots & \Delta \\
  \Delta & 0 &   &  \\
  \vdots &   & \ddots & -\Delta \\
  -\Delta &   & \Delta & 0
\end{array}
\right).
\end{equation}

With the above matrices, we can determine the excitation spectrum $E_{m}$ via solving the secular equation ${\rm det}\left[\left(M+N\right)\left(M-N\right)-E^2_{m}\right]=0$ \cite{secular}. Draw on the previous researches
where the energy gap is closed at the topological phase transition point \cite{proj_2,K3,MZM_2,cai,t6tong},
we assume that there is no exception in our model. Accordingly, the transition point $V_{c}$ can be solved by the
equation ${\rm det}\left[\left(M+N\right)\left(M-N\right)\right]=0$. Having known that ${\rm det}\left(M-N\right)
={\rm det}\left(M-N\right)^{T}={\rm det}\left(M+N\right)$, then the $V_{c}$ is further determined by this equation
${\rm det}\left(M-N\right)=0$. Eventually, we obtain the following constraint condition
\begin{equation}
\prod^{L}_{n=1}e^{i2\pi\alpha n}=\left(\frac{t+\Delta}{V}\right)^{L}.
\end{equation}
In the thermodynamic limit $L \rightarrow \infty$, $V_c$ has a real solution, and $V_{c}=t+\Delta$ (Here we have considered the condition
that $\alpha$ is approached by the ratio of two adjacent Fibonacci numbers). We have noticed that this analytic strategy has been
used in a Hermitian non-Abelian AAH model \cite{cai}. From our analytic result, the introduced non-Hermiticity compresses the
topologically non-trivial region.

\begin{figure}[H]
  \centering
  \includegraphics[width=0.5\textwidth]{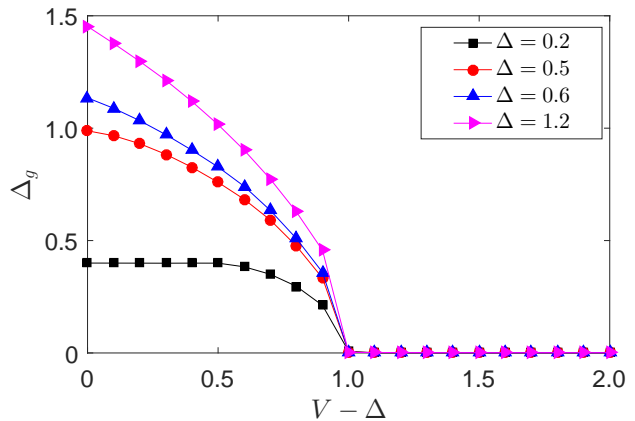}\\
  \caption{(Color online) The real energy gap $\Delta_{g}$ as a function of $V-\Delta$ under PBC. Intuitively, the gap closes at $V_c=1+\Delta$.
  The size of the system is $L=1000$.  }\label{f2}
\end{figure}

In order to verify the accuracy of the previous assumption and the analytical $V_{c}$, and to understand the
relationship between topological phase transition and gap closing, we numerically plot the variation of the real
energy gap $\Delta_{g}$ with respect to the non-Hermitian quasiperiodic potential strength $V$ under PBC, as
shown in Fig.~\ref{f2}. It is readily seen that the real energy gap closes at $V_{c}=1+\Delta$ even though the size
of the system is finite. Besides, the numerical results reflect that the assumption we made before are correct.
Moreover, the topological properties of the system are exactly protected by the real gaps, and the topological
phase transition appears with the gap closing.

In addition to the mentioned MZM and the gap-closing point, the topological phase transition is more precisely
characterized by the topological invariant $Q$. In a $p$-wave superconducting chain, the value of $Q=(-1)^\nu$
is determined by the parity of the number $\nu$ of Majorana zero modes at ends of the chain. For a periodic
invariant $p$-wave superconducting chain, Kitaev defined the topological invariant as the Pfaffian of the
Hamiltonian matrix. However, to identify the topologically non-trivial phase of a disordered superconducting
chain it is more suitable to work with the transfer matrix approach. As shown in Fig.~\ref{f1}, there appear
a pair of Majorana zero modes when $V < 1+\Delta$. We make analytical derivation of the topological phase
transition points by using the transfer matrix approach. The Hamiltonian matrix~(\ref{BDG}) can be represented in the
difference equation form
\begin{equation}
\begin{aligned}
\label{diff}
 &t(u_{n+1}+u_{n-1})+V e^{i 2\pi\alpha n} u_{n}-\Delta(v_{n+1}-v_{n-1})=Eu_{n},\\
 &\Delta(u_{n+1}-u_{n-1})-V e^{i 2\pi\alpha n} v_{n}-t(v_{n+1}+v_{n-1})=Ev_{n}.
\end{aligned}
\end{equation}
For Majorana zero modes, these equations can be represented in the transfer matrix form
\begin{equation}
 \begin{pmatrix}
 \psi_{j+1}\\
 \psi_j
\end{pmatrix}
=T_j
\begin{pmatrix}
 \psi_{j}\\
 \psi_{j-1}
\end{pmatrix}
\ \textrm{where} \ \
T_j=
\begin{pmatrix}
 \frac{V_j}{\Delta+t} & \frac{\Delta-t}{\Delta+t}\\
1 & 0
\end{pmatrix}
.
\end{equation}
If both the two eigenvalues $\lambda_1$ and $\lambda_2$ of the total transfer matrix $\textsl{T}\equiv\Pi_{j=1}^LT_j$ are
less than $1$ or larger than $1$, the system is topological non-trivial. We set $t=1$ and $\Delta>0$, then two eigenvalues $\lambda_1$ and $\lambda_2$ satisfy $|\lambda_1\lambda_2|<1$. If we set $|\lambda_1| < 1$ and $|\lambda_1| < |\lambda_2|$, the
topological property of the system is determined by the amplitude of $|\lambda_2|$. Thus, the Lyapunov exponent is defined as $R \equiv \lim_{L\rightarrow\infty}\frac{1}{L}\ln|\lambda_2(V,\Delta)|$.
\begin{figure}[H]
  \centering
  \includegraphics[width=0.45\textwidth]{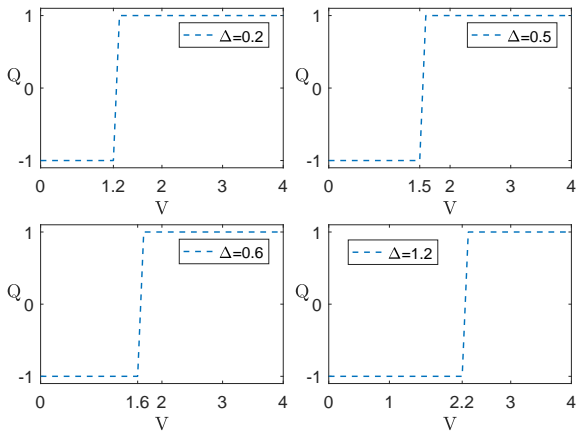}\\
  \caption{(Color online) The topological invariant $Q$ as a function of $V$ with four chosen $\Delta$. When $V<1+\Delta$, $Q=-1$,
  which corresponds to the topologically non-trivial phase; when $V>1+\Delta$, $Q=1$, which corresponds to the topologically
  trivial phase. Intuitively, $Q$ jumps at the phase transition point, i.e., the gap closing point $V_c=1+\Delta$. The size of
  the system is $L=500$.}\label{f3}
\end{figure}

For $0<\Delta< 1$, we perform a transformation $T_j=\sqrt{\xi}S\tilde{T}_jS^{-1}$ with $S={\rm diag}(\xi^{1/4},1/\xi^{1/4})$ and
$\xi=\frac{1-\Delta}{1+\Delta}$.
%The matrices
%$\tilde{T}_j=
%\left(
%\begin{matrix}
% \frac{V_j}{\sqrt{1-\Delta^2}} & -1\\
% 1 & 0
% \end{matrix}\right)$.
Thus, the total transfer matrix $\textsl{T}$ become
\begin{equation}
 \textsl{T} (V,\Delta)=(\sqrt{\frac{1-\Delta}{1+\Delta}})^LS\textsl{T}(\frac{V}{\sqrt{1-\Delta^2}}, 0)S^{-1}.
 \label{Amatrix}
\end{equation}
From the above definition and analysis, we can obtain
\begin{equation}
 R(V,\Delta)=R(\frac{V}{\sqrt{1-\Delta^2}}, 0)-\frac{1}{2}\ln(\frac{1+\Delta}{1-\Delta}).
 \label{Lyp}
\end{equation}
When $\Delta=0$, the model is reduced to the non-Hermitian AAH model \cite{mobility_2} and the Lyapunov
exponent $R(V, 0)= \ln(V)$, so $R(\frac{V}{\sqrt{1-\Delta^2}}, 0)= \ln(\frac{V}{\sqrt{1-\Delta^2}})$. According to
the above discussions, the topological transition point is at $|\lambda_2|=1$, i.e., $R(V,\Delta)=0$. From Eq.~(\ref{Lyp}),
we obtain that the topological transition point obeys $V_c = 1 +\Delta$. In Fig.~\ref{f3}, we plots the variation of
the topological invariant $Q$ versus $V$ for different $\Delta$. The topological quantum number $Q$ is evaluated by
calculating the transfer matrix numerically. The adopted numerical method is consistent with that
in Refs.~\cite{method1,method2}. When $V<1+\Delta$, $Q=-1$ which corresponds to the topologically non-trivial phase; when $V>1+\Delta$,
$Q=1$ which corresponds to the topologically trivial phase. Intuitively, $Q$ jumps at the phase transition point, i.e.,
the gap-closing point $V_{c}=1+\Delta$.

\section{Localized transition and critical states}
\label{n4}
\begin{figure}[H]
	\centering
	\includegraphics[width=0.5\textwidth]{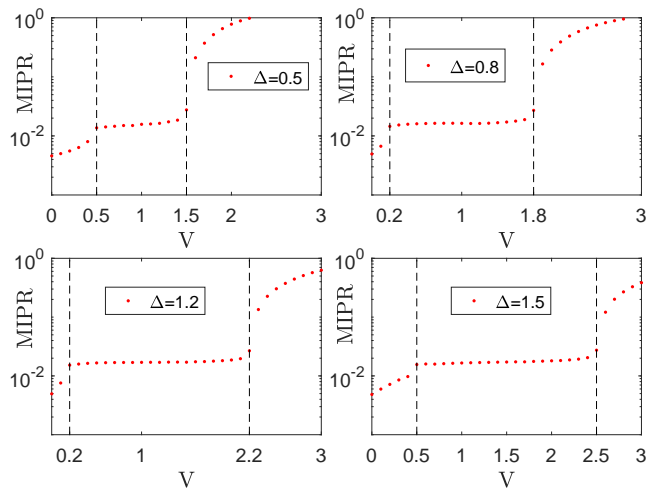}\\
	\caption{(Color online) MIPR as a function of $V$ with different $\Delta$. The dashed lines show the sharp increase
	of the MIPR at phase boundaries $V_{ec}=1-\Delta$ and $V_{c}=1+\Delta$. The total number of sites is set to be $L=500$. }
	\label{f4}
\end{figure}

Recalling the localized distributions of the lowest excitation modes in Fig.~\ref{f1} when $V>V_c$, we are aware that there is
an Anderson localization phase transition as the topological phase transition happens. Figure \ref{f4} plots the variation of MIPR
as a function of the potential strength $V$ with various $\Delta$. Intuitively, the MIPR
increases steeply at $V_{c}$ and approaches $1$. Such a phenomenon signals a delocalization-localization phase
transition, and the region where $V>V_{c}$ denotes the Anderson localization phase. However, the region where $V<V_{c}$ is not
necessarily extended phase. Instead, it is divided into two phases, i.e., the extended phase and the critical phase, whose
MIPR is greater than that of the extended phase and less than that of the localized phase, i.e., forms a platform. The extended-critical phase transition point $V_{ec}$ is readily
seen at $V_{ec}=1-\Delta$.

\begin{figure}[H]
	\centering
	\includegraphics[width=0.5\textwidth]{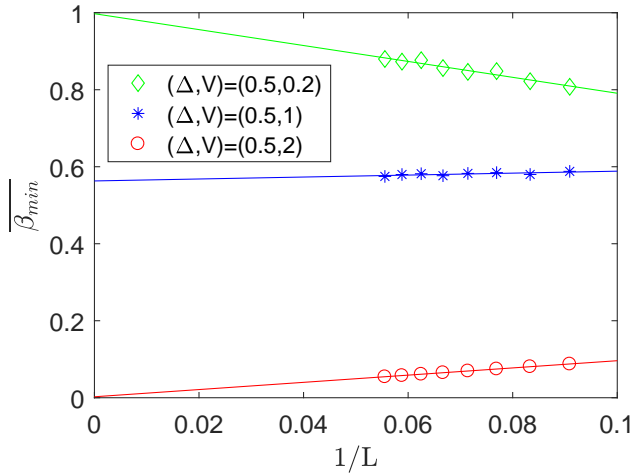}\\
	\caption{(Color online) $\overline{\beta_{min}}$ as a function of $1/L$ at $(\Delta,V)=(0.5,0.2)$, $(0.5,1)$,
	and $(0.5,2)$. These three points are located in the extended, the critical, and the localized phases, respectively. }
	\label{f5}
\end{figure}

We further validate our analysis using the fractal theory, which has been widely applied in the quasiperiodic models \cite{proj_2,t6tong,fractal_1,fractal_2,fractal_3,fractal_4}.
The size of the system $L$ is chosen as the $j$th Fibonacci number $F_{j}$. The advantage of this arrangement
is that the golden ratio can be approximately replaced by the ratio of the nearest two Fibonacci numbers, i.e.,
$\alpha=(\sqrt{5}-1)/2=\lim_{j\rightarrow\infty} F_{j-1}/F_{j}$. Then a scaling index $\beta_{m,n}$ can be extracted from the
on-site probability $P_{m,n}=u^2_{m,n}+v^2_{m,n}$ by
\begin{equation}
P_{m,n} \sim (1/F_{j})^{\beta_{m,n}}.
\end{equation}
As the fractal theorem tells, when the wave functions are extended, the maximum of $P_{m,n}$ scales as $max(P_{m,n})
\sim (1/F_{j})^1$, implying $\beta_{min}=1$. On the other hand, when wave functions are localized, $P_{m,n}$ peaks at very
few sites and nearly zero at the others, suggesting $max(P_{m,n}) \sim (1/F_j)^0$ and $\beta_{min}=0$. As for the critical
wave functions, the corresponding $\beta_{min}$ is located within the interval $\left(0,~1\right)$. For our system with $L=F_{j}$
sites, there are $2F_{j}$ eigenstates. Therefore, we can distinguish the extended, the critical, and the localized wave functions
by the average of $\beta_{min}$ (denoted by $\overline{\beta_{min}}$) over all the eigenstates , and $\overline{\beta_{min}}$
is expressed as
\begin{equation}
\overline{\beta_{min}}=\frac{1}{2L}\sum^{2L}_{m=1}\beta^{m}_{min}.
\end{equation}

Figure \ref{f5} shows the $\overline{\beta_{min}}$ as a function of $1/L$ for various parameter points $(\Delta, V)$.  We find
that $\overline{\beta_{min}}$ tends to $1$ at $(\Delta, V)=(0.5, 0.2)$ when $L$ is infinite, suggesting that the system is in the
extended phase. $\overline{\beta_{min}}$ extrapolates to zero at $(\Delta, V)=(0.5, 2)$, indicating that the system is in the
localized phase. For $(\Delta, V)=(0.5,1)$, the corresponding $\overline{\beta_{min}}$ in the thermodynamic limit is intuitively
between $0$ and $1$. We emphasize that such an analysis strategy works for other parameter points as well, and results can
be obtained accordingly. Hence, we can finally verify that there are extended and critical phases in the topologically non-trivial
region, and that the phase transition point is indeed at $V_{ec}= t-\Delta$. Meanwhile, we can also confirm that topological phase
transition is accompanied by Anderson localized phase transition, and the phase transition point is $V_c$.

\section{non-conventional real-complex transition}
\label{n5}
\begin{figure}[H]
	\centering
	\includegraphics[width=0.5\textwidth]{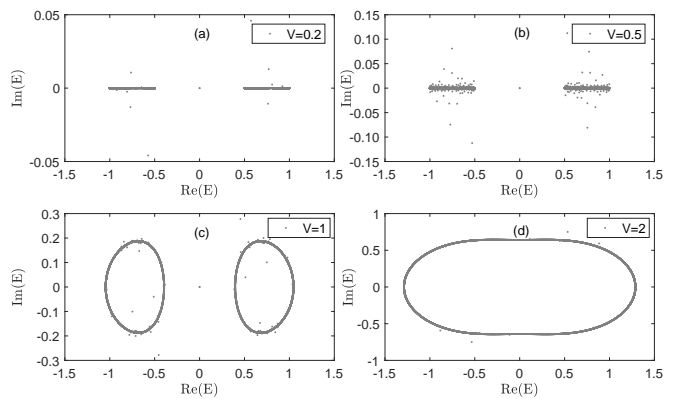}\\
	\caption{(Color online) The eigenenergies of Eq.~(\ref{eq1}) with $\Delta=0.5$ and $L=5000$ under OBC. (a) $V=0.2$ is taken
	from the extended and topologically non-trivial phase. The eigenenergies are totally real.
%, and the MZM within the real bulk gap is located at $\rm{Re(E)=0}$.
   (b) $V=0.5$ is taken at the extended-critical transition point, the imaginary parts of eigenenergies have a certain width. (c) $V=1$ is taken from the critical phase, the imaginary parts of eigenenergies are completely broadening. (d) $V=2$ is taken form the localized phase, the imaginary parts of eigenenergies are also completely broadening.}
	\label{f6}
\end{figure}

Due to the system being non-Hermitian, we turn our attention back to the energy spectrum of the system. According to precious works, the phenomenon that real-complex transition of the energy mainly exists in a class of systems with $\mathcal{PT}$ symmetry \cite{mobility_2,MZM_2,PT,PT_1,PT_2,PT_3,PT_4,PT_5,PT_6,PT_7,PT_8}.
However, for our system without $\mathcal{PT}$ symmetry, the phenomenon of real-complex transition still exists. One can say that this is a non-conventional real-complex transition, which is different from the conventional $\mathcal{PT}$ symmetric
transition. We take $\Delta=0.5$ and fix the size of the system $L=5000$. In Fig.~\ref{f6}, we display the eigenenergies of Eq.~(\ref{eq1}) with
various $V$ under OBC. As the figure shows, when $V=0.2$, the eigenenergies are real, and the system is in the extended and topologically
non-trivial phase. The finite ``bad" imaginary energies can be interpreted as the result of the finite size effect. $V=0.5$ is taken at the extended-critical transition point. Although the system is still in the topologically non-trivial phase, the imaginary parts of eigenenergies have a certain width. $V=1$ is in the critical and topologically non-trivial phase, it can be distinctly shown that the eigenenergies of the system are complex. The similar phenomenon also occurs in the case of $V=2$, in which the system is in the localized and topologically trivial phase. We have also checked other combinations of parameters and get the same results as expected. Accordingly, we settled on such a conclusion that only the extended phase support the fully real eigenenergies, providing a new result to explore the rich physics of non-Hermitian systems.

\begin{figure}%[H]
	\centering
	\includegraphics[width=0.5\textwidth]{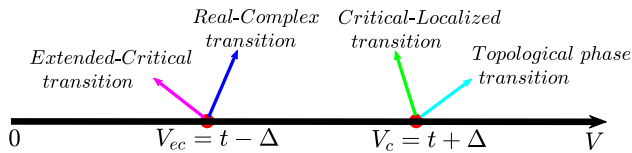}\\
	\caption{(Color online) Phase diagram of the model in this paper. $V_{ec}=t-\Delta$ (the left red dot) is the transition point of the extended-critical transition and the real-complex transition. $V_{c}=t+\Delta$ (the right red dot) is the transition point of the critical-localized transition and the topological phase transition.}
	\label{f7}
\end{figure}

Synthesizing the above analyses, we finally obtain the total phase diagram of the system, which is shown in Fig.~\ref{f7}.
As the diagram shows, the left red dot denotes the extended-critical and the real-complex transition point $V_{ec}$, satisfying $V_{ec}=t-\Delta$.
The right red dot corresponds to the critical-localized and the topological phase transition point $V_{c}$, satisfying $V_{c}=t+\Delta$.
\section{Summary}
\label{n6}

In summary, we have studied the topological properties and investigated the extended, critical and localized phases of a one-dimensional $p$-wave superconductor subject to the non-Hermitian quasiperiodic potentials.
By analysing the energy spectrum, it is shown that there are MZMs protected by the energy gap. We demonstrate that the topological phase transition is accompanied by the Anderson localization transition, and the analytic topological transition point is verified by calculating the energy gap and the topological invariant. Furthermore, we find there is a critical region separated form the extended region in the topologically non-trivial phase, and the extended-critical transition point is numerically obtained by the MIPR and the fractal analysis. Surprisingly, for our system without $\mathcal{PT}$ symmetry, we find a
non-conventional real-complex transition of the eigenenergies, and the energies in the extended phase are fully real. Unfortunately, we are failed to obtain an analytical expression of the real-complex transition point. However, it remains an open question to explore the relationship between the extended phase and the real energy, even if there is no $\mathcal{PT}$ symmetry.

%%%%%%%%%%%%%%%%
\begin{acknowledgments}
This work is supported by the National Natural Science Foundation of China (Grants No. 11674051, No. 11835011 and No. 11774316).

\end{acknowledgments}
%%%%%%%%%%%%%%%%%%%%%%%

%\begin{thebibliography}{10}

%\end{thebibliography}
\end{document}